\documentclass[12pt]{article}
\usepackage[letterpaper,margin=1in]{geometry}
\usepackage{amssymb}
\usepackage{amsmath}
\usepackage{amsfonts}
\usepackage{bm}
\usepackage{econometrics}
\usepackage{graphicx}
\usepackage{color}
\usepackage{hyperref}
\hypersetup{
   colorlinks,
   citecolor=black,
   filecolor=black,
   linkcolor=black,
   urlcolor=blue
}

\def\bi{\begin{itemize}}
\def\ei{  \end{itemize}}
\newcommand{\be}{\begin{equation}}
\newcommand{\ee}{\end{equation}}
\newcommand{\ba}{\begin{eqnarray}}
\newcommand{\ea}{\end{eqnarray}}

\newcommand{\ts}[1]{{\mbox{\rm \tiny #1}}}

\usepackage{xcolor}
\newcommand{\parens}[1]{\left(#1\right)}
\definecolor{codegreen}{rgb}{0,0.6,0}
\definecolor{codegray}{rgb}{0.5,0.5,0.5}
\definecolor{codepurple}{rgb}{0.58,0,0.82}
\definecolor{backcolour}{rgb}{0.9,0.9,0.9}

\usepackage{listings}
\usepackage[utf8]{inputenc}
\usepackage{amsmath}
\usepackage{amsfonts}
\usepackage{amssymb}
\usepackage{amsthm}
\usepackage{epsf}
\usepackage{array}
\usepackage{graphicx}
\usepackage{color}
\usepackage{bm}
\usepackage{bbm}
\usepackage{url}
\usepackage{color}
\usepackage{setspace}
\usepackage{dsfont}

\newtheorem{proposition}{Proposition}
{
\theoremstyle{definition}

}

\title{Proofs that the Gerber Statistic is Positive Semidefinite}

\author{
   S. Gerber, H. Markowitz, P.A. Ernst, Y. Miao, B. Javid, and P. Sargen
}

\begin{document}
\maketitle

\begin{abstract}
\noindent In this brief note, we prove that both forms of the Gerber statistic introduced in \cite{Gerber} are positive semi-definite.

\end{abstract}

\section{Theory}
We first recall the background theory of the Gerber statistic, as in \cite{Gerber}. Let $r_{tk}$ be the return of security $k$ at time $t,$ for $k = 1, \ldots, K$ securities and $t = 1, \ldots, T$ time periods.  For every pair $(i,j)$ of assets for each time $t$, we convert each return observation pair $(r_{ti},r_{tj})$ to a joint observation $m_{ij}(t)$ defined as:
\[
   m_{ij}(t) =
   \left\{
      \begin{array}{l}
         +1  \text{ if }
            r_{ti} \geq +H_i \text{ and } r_{tj} \geq +H_j, \\[1ex]
         +1  \text{ if }
            r_{ti} \leq -H_i \text{ and } r_{tj} \leq -H_j, \\[1ex]
         -1  \text{ if }
            r_{ti} \geq +H_i \text{ and } r_{tj} \leq -H_j, \\[1ex]
         -1  \text{ if }
            r_{ti} \leq -H_i \text{ and } r_{tj} \geq +H_j, \\[1ex]
         \phantom{+}0   \text{ otherwise,}
      \end{array}
   \right.
\]
where $H_k$ is a threshold for security $k.$  The joint observation $m_{ij}(t)$ is therefore set to $+1$ if the series $i$ and $j$ simultaneously pierce their thresholds in the same direction at time $t,$ to $-1$ if they pierce their thresholds in opposite directions at time $t$, or to zero if at least one of the series does not pierce its threshold at time $t$. We refer to a pair for which both components pierce their thresholds while moving in the same direction as a {\em concordant} pair, and to one whose components pierce their thresholds while moving in opposite directions as a {\em discordant} pair. We set the threshold $H_k$ for security $k$ to be
\[
   H_k = c \sigma_k
\]
where $c$ is some fraction (say $1/2$) and $\sigma_k$ is the sample standard deviation of the return of security $k.$  

\indent The Gerber statistic for a pair of assets is then initially defined as
\be
   \label{ogs}
   g_{ij} =
   \frac{
      \sum_{t=1}^T m_{ij}(t)
   }{
      \sum_{t=1}^T \bigl|m_{ij}(t)\bigr|
   }.
\ee
Letting $n_{ij}^c$ be the number of concordant pairs for series $i$ and $j$, and letting $n_{ij}^d$ be the number of discordant pairs, equation \eqref{ogs} is equivalent to
\[
   g_{ij} =
   \frac{
      n_{ij}^c - n_{ij}^d
   }{
      n_{ij}^c + n_{ij}^d
   }.
\]

Let us define $\mR \in \mathbb{R}^{T \times K}$ to be the return matrix having $r_{tk}$ in its $t$-th row and $k$-th column.  We may also define $\mU$ as a matrix with the same size as $\mR$ having entries $u_{tj}$ such that
\[
   u_{tj} =
   \left\{
      \begin{array}{ll}
         1 & \text{ if } r_{tj} \geq +H_j, \\[1ex]
         0 & \text{ otherwise.}
      \end{array}
   \right.
\]
With this definition, the matrix of the number of samples that exceed the upper threshold is
\[
   \mN^\ts{UU} = \mU^\top \mU.
\]
Specifically, we have the useful property that the $ij$ element $n^\ts{UU}_{ij}$ of $\mN^\ts{UU}$ is the number of samples for which both time series $i$ exceeds the upper threshold and for which time series $j$ simultaneously exceeds the upper threshold.

Similarly, define $\mD$ as the matrix with the same size as $\mR$ having entries $d_{tj}$ such that
\[
   d_{tj} =
   \left\{
      \begin{array}{ll}
         1 & \text{ if } r_{tj} \leq -H_j, \\[1ex]
         0 & \text{ otherwise.}
      \end{array}
   \right.
\]
With this definition, the matrix of the number of samples that go below the lower threshold is
\[
   \mN^\ts{DD} = \mD^\top \mD.
\]
Again, we have the useful property that the $ij$ element $n^\ts{DD}_{ij}$ of $\mN^\ts{DD}$ is the number of samples for which both time series $i$ goes below the lower threshold and for which time series $j$ simultaneously goes below the lower threshold. The matrix containing the numbers of concordant pairs is therefore
\[
   \mN_\ts{CONC} = \mN^\ts{UU} + \mN^\ts{DD} = \mU^\top \mU + \mD^\top \mD.
\]
It can be shown that the matrix containing the numbers of discordant pairs is
\[
   \mN_\ts{DISC} = \mU^\top \mD + \mD^\top \mU.
\]
We can now write the Gerber matrix $\mG$ (i.e., the matrix that contains $g_{ij}$ in its $i$-th row and $j$-th column) in the equivalent matrix form
\[
   \mG =
   \left(
      \mN_\ts{CONC} - \mN_\ts{DISC}
   \right)
   \oslash
   \left(
      \mN_\ts{CONC} + \mN_\ts{DISC}
   \right)
\]
where the symbol $\oslash$ represents the Hadamard (elementwise) division.

\subsection{A Graphical Relationship}

To simplify the description of various versions of the Gerber statistic, it is useful to consider the following graphical representation for the relationship between two securities:
\begin{center}
   \begin{tabular}{ccc}
      $UD$ & $UN$ & $UU$ \\
      $ND$ & $NN$ & $NU$ \\
      $DD$ & $DN$ & $DU$
   \end{tabular}
\end{center}
Let the rows here represent categorizations of security $i,$ and let the columns represent categorizations of security $j.$  The boundaries between the rows and the columns are the chosen thresholds.
The letter $U$ represents the case in which a security's return lies above the upper threshold (i.e., is up).
The letter $N$ represents the case in which a security's return lies between the upper and lower thresholds (i.e., is neutral).
The letter $D$ represents the case in which a security's return lies below the lower threshold (i.e., is down).

For example, if at time $t,$ the return of security $i$ is above the upper threshold, this observation lies in the top row.  If, at the same time $t,$ the return of security $j$ lies between the two thresholds, this observation lies in the middle column.  Therefore, this observation lies in the UN region. Over the history, $t = 1, \ldots, T,$ there will be observations scattered over the nine regions.  Let $n_{ij}^{pq}$ be the number of observations for which the returns of securities $i$ and $j$ lie in regions $p$ and $q,$ respectively, for $p, q \in \{U, N, D\}$.  

The covariance matrix constructed from the ``initial'' Gerber statistic as defined in equation \eqref{ogs} was often not positive semidefinite (PSD).  We therefore tested some alternative versions that were PSD.  Two alternative versions are as follows. It is important to note that in \cite{Gerber}, we employed ``Gerber Statistic 2.'' \\

\noindent \textbf{Gerber Statistic 1}: 
Let the matrix $\mG^{(1)}$ be defined as containing $g_{ij}^{(1)}$ in its $i$th row and $j$th column, with
\begin{equation}\label{GS2}
   g_{ij}^{(1)} =
   \frac{
      n_{ij}^\ts{UU} + n_{ij}^\ts{DD} - n_{ij}^\ts{UD} - n_{ij}^\ts{DU}
   }{
      \sqrt{n_{ij}^\ts{(A)}  n_{ij}^\ts{(B)}}
   }
\end{equation}
where
\begin{align*}
   n_{ij}^{(A)} & = n_{ij}^\ts{UU} + n_{ij}^\ts{UN} + n_{ij}^\ts{UD} +
                    n_{ij}^\ts{DU} + n_{ij}^\ts{DN} + n_{ij}^\ts{DD} \\
   n_{ij}^{(B)} & = n_{ij}^\ts{UU} + n_{ij}^\ts{NU} + n_{ij}^\ts{UD} +
                    n_{ij}^\ts{UD} + n_{ij}^\ts{ND} + n_{ij}^\ts{DD}.
\end{align*}
\noindent \textbf{Gerber Statistic 2}: 
Let the matrix $\mG^{(2)}$ be defined as containing $g_{ij}^{(2)}$ in its $i$th row and $j$th column, with
Let us consider
\[
   g_{ij}^{(2)} =
   \frac
      {\sum_{t=1}^T m_{ij}(t)}
      {T- n_{ij}^{NN}}.
\]
This can be written in terms of the above alternative notation as\\\\
\begin{equation}\label{GS1}
   g_{ij}^{(2)} =
   \frac{
      n_{ij}^\ts{UU} + n_{ij}^\ts{DD} - n_{ij}^\ts{UD} - n_{ij}^\ts{DU}
   }{
      {T - n_{ij}^\ts{NN}}
   }.
\end{equation}

Let $\mH = \mN_\ts{CONC} - \mN_\ts{DISC},$ and let $\vh = \sqrt{\diag(\mH)}$ be the vector of square roots of the diagonal elements of $\mH$ (which are all positive).  We now see that the first alternative version of the Gerber statistic can be written in the matrix form
\[
   \mG^{(1)} = (\mN_\ts{CONC} - \mN_\ts{DISC}) \oslash (\vh \vh^\top).
\]
Written differently, letting $\mJ = \mJ^\top$ be the diagonal matrix with the inverse of the $i$-th element of $\vh$ in its $i$-th diagonal position, we have
\[
   \mG^{(1)} = \mJ^\top (\mN_\ts{CONC} - \mN_\ts{DISC}) \mJ.
\]

Portfolio optimizers require the covariance matrix of securities' returns to be positive semidefinite. We therefore prove that both matrices $\mG^{(1)}$ and $\mG^{(2)}$  are positive semidefinite.

\section{Gerber Statistic 1}

First, it is useful to note that the numerator matrices of both $\mG^{(1)}$  and $\mG^{(2)}$ are positive semidefinite. We can see this as follows. From the definitions of $\mN_\ts{CONC}$ and $\mN_\ts{DISC},$ the numerator matrix can be written in the following squared form:
\begin{align*}
   \mH & = \mN_\ts{CONC} - \mN_\ts{DISC} \\
       & = \mU^\top \mU + \mD^\top \mD - \mU^\top \mD - \mD^\top \mU \\
       & = (\mU - \mD)^\top (\mU - \mD).
\end{align*}
Therefore, for arbitrary but non-zero $\vx,$
\[
   \vx^\top \mH \vx = \vx^\top  (\mU - \mD)^\top (\mU - \mD) \vx = \vu^\top \vu \geq 0.
\]

We now prove that the matrix $\mG^{(1)}$ is positive semidefinite.
\begin{proposition}
The matrix $\mG^{(1)}$ is positive semidefinite.
\end{proposition}
\begin{proof} 
We write
\begin{align*}
   \vx^\top \mG^{(1)} \vx & = \vx^\top \mJ^\top \mH \mJ \vx \\
                          & = \vx^\top \mJ^\top (\mU - \mD)^\top (\mU - \mD) \mJ \vx = \vu^\top \vu \geq 0.
\end{align*}
This concludes the proof.
\end{proof}

\section{Gerber Statistic 2}
We now prove that the matrix $\mG^{(2)}$  is positive semidefinite. This will be eased by defining a new matrix $\mF$,  with entires $f_{tj}$ such that
\[
   f_{tj} =
   \left\{
      \begin{array}{ll}
         1 & \text{ if } r_{tj} \geq +T_j, \\[1ex]
         -1 & \text{ if } r_{tj} \leq -T_j, \\[1ex]
         0 & \text{ otherwise.}
      \end{array}
   \right.
\]
Let $\langle F_i,F_j \rangle$ be the dot product of columns $i$ and $j$ of matrix $\mF$ corresponding to assets $i$ and $j$. This leads to the following proposition. 

\begin{proposition}
$\mG^{(2)}$ is positive semidefinite. 
\end{proposition}

\begin{proof}
We first note that the numerator of $\mG^{(2)}$ can be expressed as $\mF^T \mF$. It is therefore positive semidefinite. 
We now consider the denominator of $\mG^{(2)}$. Let us define the matrix $\mX$ with entires $$x_{i,j}:=\frac{n_{ij}^\ts{NN}}{T}.$$ The entries $g_{i,j}$ in the denominator of the matrix $\mG^{(2)}$ become $T(1-x_{i,j})$. Noting that all entries $0\leq x_{i,j}<1$, (the case $x_{i,j}=1$ would correspond to the case where all observations fall into the $NN$ region, which we exclude) we can  rewrite the entires 
$$g_{i,j}=\frac{\langle F_i,F_j \rangle}{T(1-x_{i,j})}=\frac{\langle F_i,F_j \rangle}{T}(1+x_{i,j}+x_{i,j}^2+x_{i,j}^3+...).$$
We now show that the matrix $\mX$ is positive semidefinite. To see why, let us perform another transformation of returns. Consider the new matrix $\mP$ with entires $p_{i,j}$ as
$$p_{i,j}=1-|f_{t,j}|.$$
Therefore,
$$x_{i,j}:=\frac{n_{ij}^\ts{NN}}{T}=\frac{\langle P_i,P_j \rangle}{T},$$
where $\langle P_i,P_j \rangle$ is the dot product of columns $i$ and $j$ of the matrix $\mP$ corresponding to assets $i$ and $j$. Hence, $$\mX=\frac{\mP^T\mP}{T},$$ and is therefore positive semidefinite. We can thus write 
$$g_{i,j}=\frac{\langle F_i,F_j \rangle}{T}\parens{1+\frac{\langle P_i,P_j \rangle}{T}+\frac{\langle P_i,P_j \rangle}{T}\frac{\langle P_i,P_j \rangle}{T}+\frac{\langle P_i,P_j \rangle}{T}\frac{\langle P_i,P_j \rangle}{T}\frac{\langle P_i,P_j \rangle}{T}+...}$$
Using the elementwise product $\otimes$, we can now rewrite the matrix $\mG^{(2)}$  as

$$\mG^{(2)}=\frac{\mF^T\mF}{T}\otimes\parens{\mathbf{1}+\frac{\mP^T\mP}{T}+\parens{\frac{\mP^T\mP}{T}}\otimes \frac{\mP^T\mP}{T}+\parens{\frac{\mP^T\mP}{T}\otimes \frac{\mP^T\mP}{T}\otimes \frac{\mP^T\mP}{T}}+...}$$
By the Schur product lemma, since each of the terms of $\mG^{(2)}$ is positive semidefinite,  $\mG^{(2)}$ must also be positive semidefinite. This finishes the proof.

\end{proof}


\begin{thebibliography}{99}




\bibitem{Gerber}  Gerber, S., Markowitz, H.M., Ernst, P.A., Miao, Y , Javid, B., and Sargen, P. (2022) The Gerber statistic: a robust co-movement measure for portfolio optimization. \textit{The Journal of Portfolio Management}, 48(3): 87-102




\end{thebibliography}
\end{document}